# Mitigation of Delayed Management Costs in Transaction-Oriented Systems

*Dmitry Zinoviev+, Dan Stefanescu+, Hamid Benbrahim\*, Greta Meszoely+*

+Suffolk University, Boston
\*TD Ameritrade

**Abstract**
Abundant examples of complex transaction-oriented networks (TONs) can be found in a variety of disciplines, including information and communication technology, finances, commodity trading, and real estate. A transaction in a TON is executed as a sequence of subtransactions associated with the network nodes, and is committed if every subtransaction is committed. A subtransaction incurs a two-fold overhead on the host node: the fixed transient operational cost and the cost of long-term management (e.g. archiving and support) that potentially grows exponentially with the transaction length. If the overall cost exceeds the node capacity, the node fails and all subtransaction incident to the node, and their parent distributed transactions, are aborted. A TON resilience can be measured in terms of either external workloads or intrinsic node fault rates that cause the TON to partially or fully choke. We demonstrate that under certain conditions, these two measures are equivalent. We further show that the exponential growth of the long-term management costs can be mitigated by adjusting the effective operational cost: in other words, that the future maintenance costs could be absorbed into the transient operational costs.

# Introduction

Distributed transactional computer and communication systems have been extensively studied over the course of the past two decades [1, 2]. Research ranges in scope and size from compact storage area networks [3] to large-scale distributed databases [4]. Since then, the concept of a transaction-oriented network has been generalized to embrace financial [5], supply [6], trade [7], trust [8], and even social networks. Transaction-oriented networks provide atomic, consistent, isolated, and durable execution of operations. They support scalability, fault tolerance, reliability, and predictability [9]. Unfortunately, the distributed nature of TONs also makes them vulnerable to targeted and random attacks [10, 11].

An inevitable attribute of a distributed transaction is its execution cost (which could be expressed in terms of time, electric power, produced heat, CPU cycles, memory/storage requirements, manpower or even directly in monetary amounts). The costs are associated both with executing and committing (or aborting) a transaction and with delayed maintenance costs (such as auditing, archiving, warranties, and other legal support). High execution and management costs may, on one hand, make transactions prohibitively expensive and, on the other hand, overload and even fully disable (choke) the network.

The goal of this paper is to study the effect of both aspects of transaction costs on the overall network performance. For this study, we chose a stylized version of a network where nodes process generic transactions requiring certain capacity and processing time. We particularly focus on the possibility of converting the delayed costs to the direct upfront costs, since such conversion would greatly simplify further network analysis.

Our results have been partially presented at the High Performance Computing Symposium, San Diego, CA [12]. The new contribution of this paper includes:

1. the introduction of a new transaction cost model that has a one-time upfront and a recurring long-term component;
2. the introduction of a mechanism for transforming the long-term costs into upfront direct costs.

The paper is organized as follows: in the first section, we describe the new network model; in the second section, we present a summary of previously published results from [12] and present and discuss the new simulation results; finally, in the third section we propose a cost flattening mechanism.

## Model Overview

In this paper, we simulate and discuss a random Erdős-Rényi transaction-oriented network [13] that consists of *N*=1000 identical nodes connected by $d\,N\,(N-1)$ links (that is, the network density is *d*). Each network node represents an entity that can simultaneously execute up to *C* independent subtransactions. Each subtransaction takes time $\tau_0$ to complete (the time does not depend on the total load on the node). The network is simulated for the duration of S=36,500 time steps.

Each node is also a potential transaction source and sink. During the simulation, transactions are injected uniformly across the network. The delays between subsequent transactions are drawn from the exponential distribution *E*(1/*r*), where *r* is the mean injection rate.

All transactions injected in the network are distributed. A master transaction T consists of L=10 subtransactions $T_i$ (i=1...L). A master transaction is committed if all its subtransactions are committed. Alternatively, the transaction is aborted.

Transaction routing uses an opportunistic routing strategy: the node for the next subtransaction is chosen uniformly at random from all neighbors of the current node. If the next node is disabled, then another neighbor is chosen. It is possible for the next subtransaction $T_{i+1}$ to be executed by the same node as the previous subtransaction $T_{i-1}$ (but not as the current subtransaction $T_i$). If all neighbors are disabled, the subtransaction and the corresponding master transaction are aborted.

A network nodes can become disabled either by overloading (when the actual load at a node reaches or exceeds its capacity *C*) or by random internal faults that are injected after an initial delay drawn from the exponential distribution $E(T_f)$. A node shutdown in a computer or communication network or in a distributed database can be caused, e.g., by overheating [14] or thrashing (excessive swapping to the secondary storage [15]), and in a financial network, a shutdown may correspond to or be a consequence of a bankruptcy procedure [16].

Initially, all nodes in a network are alive and can perform their tasks. Once disabled, however, a node is not restarted and remains disabled for the rest of the simulation run (we speculate that recovery may not be feasible or even possible, especially in  in autonomous unmanaged networks).

There is a cost model associated with our distributed transactions. A subtransaction incurs a two-fold overhead on the host node of every subtransaction: the fixed transient operational cost $\Psi_0$ and the long-term management (e.g., archiving and support) overhead $\Delta\Psi=\Psi_0(\alpha^{i-1}-1)$, where $i=1..L$ is the subtransaction index and α is a long-term impact factor. The case of α=1 corresponds to no long-term management costs, and the case of α<1 corresponds to discounted execution. The total cost of a committed transaction is, therefore, given by the following equation:

$$\Psi = \Sigma_{i=1}^{L}\Psi_0\alpha^i = \Psi_0\frac{\alpha^L-1}{\alpha-1}. \tag{1}$$

The overall incurred overhead at every node continuously and exponentially decays over time. In our simulations, we calculate the decay using a discrete approximate equation:

$$\Xi(Q,t) = \Xi(Q,t_D)\, e^{-(t-t_S)/H}. \tag{2}$$

Here, $t$ is the current time, $t_D$ is the time when the decay was calculated most recently, $\Xi(Q,t)$ is the cumulative cost of all subtransactions at node $Q$ at time $t$, and $H$ is the decay time. For most calculations presented in this paper, $H=30\,\tau_0$.

An alternative interpretation of the transaction cost is a risk associated with the transaction. This interpretation is especially vivid for financial and real estate networks.

The network simulator has been implemented in C++ using a discrete event simulation package developed at the Mathematics and Computer Science Department of Suffolk University.

## Simulation Results

The goal of the simulation was two-fold:

1. To study the TON's resilience with respect to external and internal factors;
2. To investigate if and how long-term transaction management costs can be converted to one-time upfront transaction execution costs.

Some of the following results were presented at the High Performance Computing Symposium, San Diego, CA [12]. A transaction-oriented network can be failed externally, by overloading, and internally, by randomly or systematically failing the individual nodes. We simulated external overloading by starting with a fully functional network and gradually increasing the transaction injection rate r from 0 to $r_0$ until at least $10^{-6}$ of all injected transactions end up aborted. Borrowing the terminology from superconductor digital microelectronics [17], we say that the network in a *superconductive* (almost fully conductive) mode when r<$r_0$. As we continue increasing r, the fraction of the aborted transactions monotonically increase, until at some rate $r_1$ the network chokes and transgresses from the *resistive* (barely conductive) mode to the *dielectric* (non-conductive) mode. We define $\rho_0=r_0/r_1$.

Alternatively, a network may choke because too many of its internal nodes choke. To simulate internal failures, we start with a fully functional network at the border between the superconductive and resistive modes (r=$r_0$). At the fixed injection rate, we start failing random nodes after random delays (a failed node is never repaired). Naturally, this brings the network to the dielectric mode once again. We define $m_0$ to be the smallest fraction of failed nodes that cause the network to choke. Finally, the network can be brought to a halt by simultaneously increasing the injection rate from 0 onward and failing random nodes.

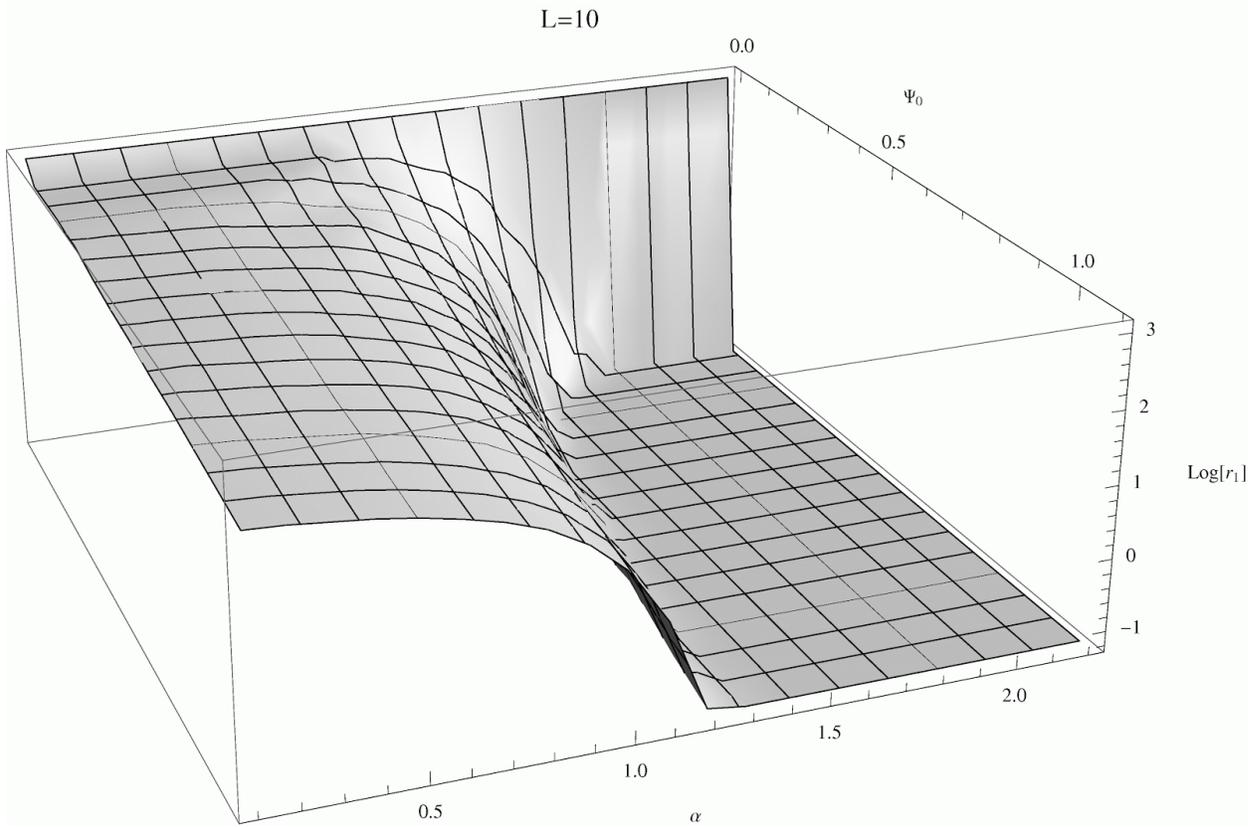

*Figure 1. Resistive-to-dielectric critical transaction injection rate $r_1$ vs the transaction cost parameters α and $\Psi_0$*

We define two transaction-oriented networks to be behaviorally equivalent if they have the same values of $r_0$, $r_1$, and possibly $m_0$. Two behaviorally equivalent networks start failing and fail under similar external conditions. We determined that there is a very strong correlation, bordering a functional dependency, between $r_{0/1}$ and node capacity C and network density d:

$$r_{0/1} \approx A_{0/1}(C-2)^{\beta_{0/1}}. \tag{3}$$

The exponents $\beta_{0/1}$ for the dense networks are ~2, and they tend to 1 as d tends to 0. Perhaps a more unexpected result is that the relationship between the network resilience parameters $m_0$ and $\rho_0$ is almost linear, with the slope of -1. That is, tolerating additional superconductive traffic $\Delta\rho_0$ is almost equivalent to disabling extra network nodes $\Delta m_0$ due to the internal faults:

$$\Delta\rho \approx -\Delta m_0. \tag{4}$$

Moreover, for a fixed node capacity C and transaction length L, there is a very strong correlation, bordering a functional dependency, between $r_0$, $r_1$, and $m_0$:

$$r_0 \simeq a\left(\sqrt{b^2+r_1^2}-b\right), \tag{5}$$

$$m_0 \simeq \Delta m - (\Delta m - 1)\sqrt{1+(\rho_0/\lambda)^2}. \tag{6}$$

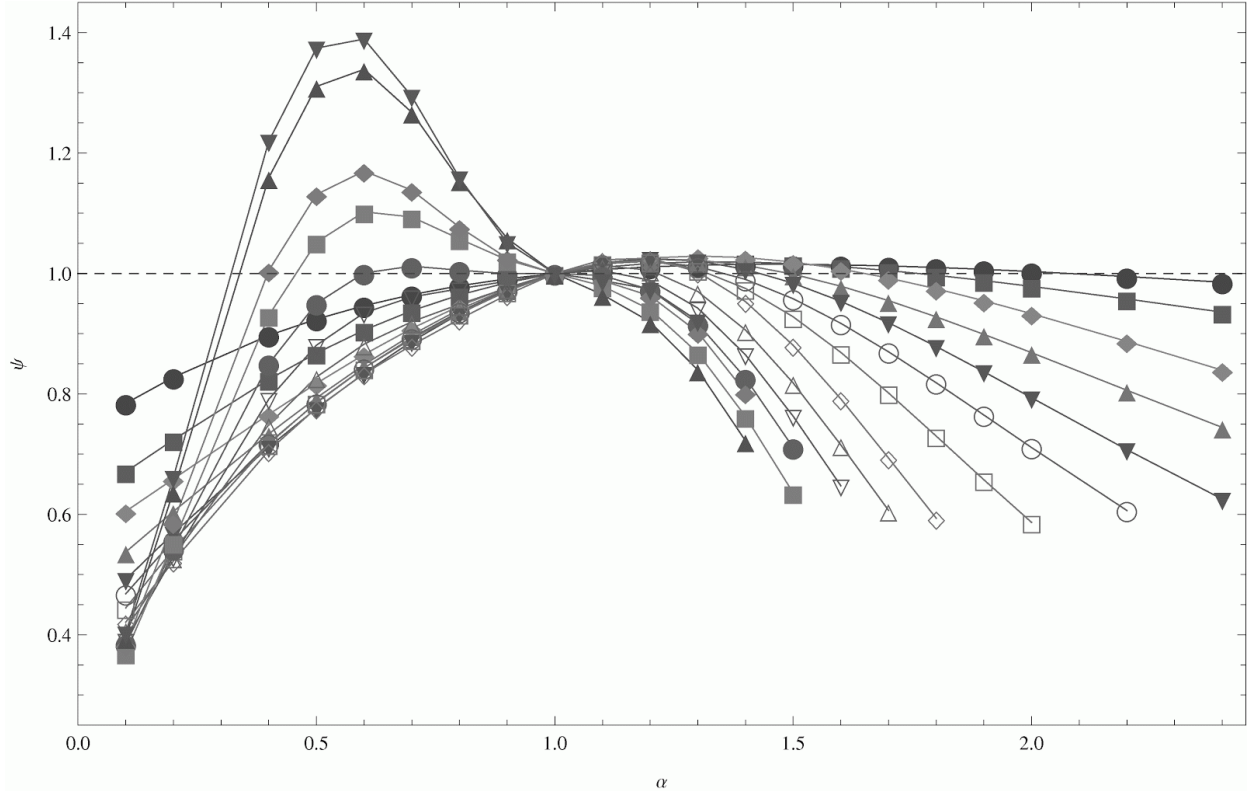

*Figure 2. Relative cost of transaction flattening for different transaction lengths as a function of α.*

Here, a≈1, b≈3...7, Δm≈1...1.5, and λ≈0.1...0.4 are fitting parameters that are slow and almost monotonous functions of L. We can, therefore, use only one of the three network resilience measures to establish behavioral equivalence between two networks that have the same configuration in terms of density d, node capacity C, and transaction length L. In other words, if one knows $r_1$, then $r_0$ and $m_0$ can be reasonably faithfully estimated from the equations (5) and (6).

Having established $r_1$ as the sole measure of TON resilience, we proceed with simulating the network with fixed d, L, and C, but with different transaction cost parameters $\Psi_0$ (operational transaction cost) and α (long-term impact factor).

## Flattening Transaction Costs

Figure 1 shows the logarithm of the resistive-to-dielectric critical injection rate $r_1$ as a function of the transaction cost parameters $\Psi_0$ and α for a TON with L=10, C=10, and d=0.5. Since $r_1$ is limiting the performance of the network, it is beneficial to keep it as high as possible. The higher values of $r_1$ correspond to lower execution costs $\Psi_0$ and lower values of the long-term impact factor α. Interestingly, the Figure 1 shows that the long-term attenuation of transaction management costs (α<1) is good for the network and results in higher $r_1$ almost regardless of the initial execution cost $\Psi_0$.

Any horizontal cross-section of the surface in Figure 1 represents a family of behaviorally equivalent TONs, that is, of networks that sustain (almost) identical transactional traffic and differ only in the cost parameters. We speculate that the equation of the cross-section in the α-$\Psi_0$ coordinate system is given by the following equation:

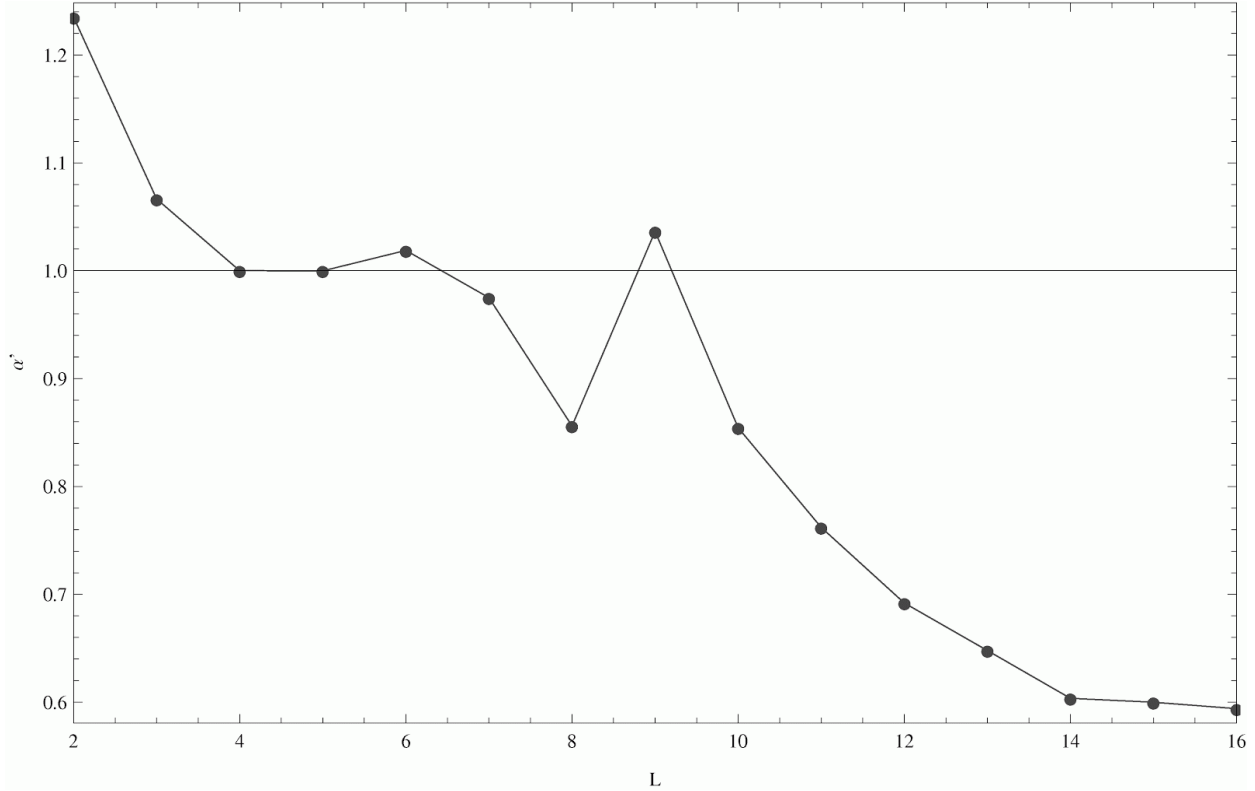

*Figure 3. Prime long-term impact factor α' vs transaction length L*

$$ln(r_1) \simeq A\, (\Psi_0)^{B_\Psi}\left((\alpha+\delta_\alpha)^2+\gamma_\alpha^2\right)^{B_\alpha}+c\ . \tag{7}$$

Eq. (7) represents a generalized hyperbola. For fixed C and d and for realistically long transactions (L>3), some fitting parameters from the equation are almost constant in L: $B_\Psi \approx 0.9$, $\gamma_\alpha \approx 1$, and $\delta_\alpha \approx -0.4$ ($\chi^2 > 0.965$ for all L between 2 and 16). The other three parameters either monotonically decrease with L ($c \approx 5.5 \rightarrow 1.3$) or monotonically increase with L ($A \approx -1.8 \rightarrow -0.0002$, $B_\alpha \approx 1 \rightarrow 15$). This is consistent with the common sense observation that for longer transactions, the long-term management costs have higher impact on the network performance than for the short transactions.

An equation for $r_0$ differs from Eq. (7) only in the values of the fitting parameters.

The case of α=1 correspond to a "flat" transaction, that is, a transaction with no long-term management costs (or discounts). For a "flat" transaction, for any feasible injection rate $ln(r_1) \leq c$, Eq. (7) can be solved with respect to $\Psi_0$:

$$\Psi_0' = \sqrt[B_\Psi]{\frac{c-ln(r_1)}{-A\left((1+\delta_\alpha)^2+\gamma_\alpha^2\right)^{B_\alpha}}}\ . \tag{8}$$

Here, $\Psi_0'$ is the efficient upfront "flat" transaction execution cost that yields a TON behaviorally equivalent to the original TON, but with no additional long-term management costs. We suggest that via Eq. (8), any TON (α, $\Psi_0$) with progressively surcharged or discounted subtransactions can be transformed into a normalized TON (1, $\Psi_0'$) with a flat subtransaction cost.

## The Price of Flattening

Let ψ be the relative cost of transaction flattening, that is, the ratio of the original transaction cost Ψ, given by Eq. (1), and the cost of the corresponding normalized transaction LΨ$_0$' from Eq. (8):

$$\psi = \frac{\Psi_0(\alpha^L - 1)}{\Psi_0'(\alpha - 1)L}. \tag{9}$$

For a transaction with no long-term costs (α=1), the flattening cost ratio is ψ=1. The flattening process reduces transaction costs only if ψ>1. Figure 2 shows the dependence of ψ on α for a network with fixed C=10 and d=0.5 and various transactions lengths L. We observe that each curve corresponding to a network with a certain transaction length has a peak above 1, but the peak is either significant and on the left of α=1 (for longer transactions L≥12) or barely visible and on the right of α=1 (for shorter transactions L≤11). In other words, it is somewhat profitable to flatten the costs of shorter transactions with long-term discounts, barely profitable to flatten shorter transactions with long-term expenses, and not profitable to flatten the costs of all other transactions. We observe that the position of the peak α' is an equivalent of prime rate [18], because the corresponding cost attenuation/amplification ensures the best cost flattening gain (Figure 3).

## Conclusion

We modeled and simulated a homogeneous, random Erdős-Rényi transaction-oriented network (TON) of general-purpose processing nodes with opportunistic routing. Every subtransaction of a distributed transaction has a constant execution (processing) cost and a progressively increasing/decreasing long-term management (e.g., archival or legal support) cost. A network node has a finite capacity: when the aggregate cost of all incident subtransactions exceeds the capacity, the node terminates. A node can also permanently fail due to an internal error. We assume that an external network user is interested, in the first place, in the critical transaction injection rates that (a) cause at least some transactions to fail over the simulation run or (b) cause the network to choke over the simulation run, and (c) in the critical internal node failure rate that causes the network to choke over the simulation run. We demonstrated that under this assumption, the network cost model can be adjusted, without affecting the resilience measures (a)-(c), to nullify long-term cost attenuation/amplification and absorb the progressive management costs into flat upfront processing costs. The flattening procedure can be used to reduce the intrinsic TON complexity, streamline TON operations, and simplify reasoning about the TON (including its mathematical and computational models).